\documentclass[a4paper,11pt]{article}
\usepackage{jheppub}
\usepackage{CJK}
\usepackage{tikz}
\usepackage{float,graphbox,makecell,tablefootnote}
\usepackage{longtable}
\usepackage{multirow}
\allowdisplaybreaks[4]
\usepackage{diagbox}
\usepackage{tabularx}
\usepackage{makecell}
\usepackage{physics}
\usepackage{alltt}
\usepackage{array}

\usepackage{amssymb} 
\usepackage{graphicx} 
\usepackage{multicol} 
\usepackage{booktabs}
\usepackage{bigstrut}
\usepackage{mathtools}
\usepackage{mathrsfs}
\usepackage{float}
\usepackage{subfig}
\usepackage{colortbl} 
\usepackage{amsthm}

\usepackage{upgreek}

\definecolor{darkred}{RGB}{173,34,48}
\definecolor{lightgreen}{rgb}{0.56, 0.69, 0.19}
\definecolor{lightblue}{rgb}{0.36, 0.51, 0.71}
\definecolor{lightyellow}{rgb}{0.88, 0.61, 0.14}
\definecolor{darkgreen}{rgb}{0.6, 0.6, 0.35}
\definecolor{lightred}{rgb}{0.99, 0.36, 0.02}
\definecolor{box1}{rgb}{0.46, 0.6, 0.45}
\definecolor{box2}{rgb}{0.62, 0.56, 0.43}
\definecolor{box3}{rgb}{0.72, 0.65, 0.17}
\newcommand{\ab}[1]{\langle #1\rangle}
\usepackage[hang]{footmisc}
\makeatletter
\ifFN@para
\else
  \long\def\@makefntext#1{%
    \ifFN@hangfoot
      \bgroup
      \setbox\@tempboxa\hbox{%
        \ifdim\footnotemargin>0pt
          \hb@xt@\footnotemargin{\@makefnmark\hss}%
        \else
          \@makefnmark\hskip-\footnotemargin      
        \fi
      }%
      \leftmargin\wd\@tempboxa
      \rightmargin\z@
      \linewidth \columnwidth
      \advance \linewidth -\leftmargin
      \parshape \@ne \leftmargin \linewidth
      \footnotesize
      \@setpar{{\@@par}}%
      \leavevmode
      \llap{\box\@tempboxa}%
      \parskip\hangfootparskip\relax
      \parindent\hangfootparindent\relax
    \else
      \parindent1em
      \noindent
      \ifdim\footnotemargin>\z@
        \hb@xt@ \footnotemargin{\hss\@makefnmark}%
      \else
        \ifdim\footnotemargin=\z@
          \llap{\@makefnmark}%
        \else
          \llap{\hb@xt@ -\footnotemargin{\@makefnmark\hss}}%
        \fi
      \fi
    \fi
    \footnotelayout#1%
    \ifFN@hangfoot
      \par\egroup
    \fi
  }
\fi
\makeatother

\usepackage{slashed}

\usepackage{tikz}

\graphicspath{{images/}}
\begin{document}

\begin{CJK*}{UTF8}{}
\CJKfamily{gbsn}

\title{Loops and legs: ABJM amplitudes from $f$-graphs}

\author[a,b,c,d]{Song He,}\emailAdd{songhe@itp.ac.cn}
\author[a,d,e]{Yao{-}Qi Zhang}\emailAdd{y.zhang59@herts.ac.uk}

\affiliation[a]{Institute of Theoretical Physics, Chinese Academy of Sciences, Beijing 100190, China}
\affiliation[b]{School of Fundamental Physics and Mathematical Sciences, Hangzhou Institute for Advanced Study and ICTP-AP, UCAS, Hangzhou 310024, China}
\affiliation[c]{Peng Huanwu Center for Fundamental Theory, Hefei 230026, China}
\affiliation[d]{School of Physical Sciences, University of Chinese Academy of Sciences, No.19A Yuquan Road, Beijing 100049, China}
\affiliation[e]{Department of Physics, Astronomy and Mathematics,
University of Hertfordshire,
Hatfield, Hertfordshire, AL10 9AB, United Kingdom}
\abstract{We initiate a systematic study on how to extract planar integrands of (supersymmetric) scattering amplitudes with $L$ loops and $n$ legs in Aharony-Bergman-Jafferis-Maldacena (ABJM) theory from the recently proposed (bosonic) generating function for squared amplitudes with $N:=n{+}L$ dual points; the latter enjoys a hidden permutation symmetry $S_N$ and is given by a linear combination of weight-$3$ planar $f$-graphs that can be recast as bipartite graphs, which manifest important properties of ABJM amplitudes. We provide evidence that it contains sufficient information to reconstruct individual amplitudes, despite the absence of squared amplitudes at odd loops. The extraction of the four-point amplitude is already non-trivial and closely parallels the extraction of five-point amplitudes in ${\cal N}=4$ super Yang-Mills (SYM) from weight-$4$ $f$-graphs: we comment on this similarity and provide new results for $n=4$ ABJM loop integrand up to $L=6$. For higher multiplicities, based on Yangian invariants (including BCFW building blocks for tree amplitudes) and an appropriate basis of planar dual conformal invariant(DCI) integrands, we disentangle six-point integrands up to two loops and eight-point tree amplitude from the squared amplitudes. Our results suggest that ABJM amplitudes of arbitrary multiplicity and loop order can be reconstructed from squared amplitudes, closely paralleling the role of $f$-graphs in $\mathcal{N}=4$ SYM.}

\maketitle
\end{CJK*}


\section{Introduction}



Recent years have witnessed remarkable progress in understanding scattering amplitudes in Quantum Field Theory, particularly in superconformal theories like four-dimensional ${\cal N}=4$ super-Yang-Mills (SYM) and three-dimensional ${\cal N}=6$ Chern-Simons-matter theory known as Aharony-Bergman-Jafferis-Maldacena (ABJM)~\cite{Aharony:2008ug,Hosomichi:2008jb}. In both theories, new mathematical structures in the planar integrands of all-loop amplitudes have been discovered, which share important hidden symmetry such as the Yangian symmetry~\cite{Drummond:2008vq,Brandhuber:2008pf,Drummond:2009fd,Bargheer:2010hn,Huang:2010qy,Gang:2010gy}, deep connections to the positive Grassmannian~\cite{Arkani-Hamed:2012zlh,Lee:2010du,Huang:2013owa,Huang:2014xza} and the remarkable geometry known as the amplituhedra~\cite{Arkani-Hamed:2013jha,Arkani-Hamed:2021iya,Damgaard:2019ztj,He:2021llb,Huang:2021jlh,He:2022cup,He:2023rou}. A long-standing open question is whether the amplitude/Wilson loop/correlator triality, analogous to the well-established SYM case~\cite{Caron-Huot:2010ryg,Alday:2010zy,Eden:2011yp,Eden:2011ku,Heslop:2022xgp}, also exists in ABJM theory ({\it c.f.} ~\cite{Henn:2010ps,Bianchi:2011rn,Bianchi:2011dg,Bargheer:2012cp,Rosso:2014oha,Berkovits:2008ic,Colgain:2016gdj}). A significant step in this direction was the recent discovery of a hidden permutation symmetry that unifies amplitudes with different multiplicities and loop orders under the constraint N=n+L~\cite{He:2025zen}, offering a powerful new framework for studying the squared amplitudes, or cross-section, as well as (loop integrands of) scattering amplitudes themselves in ABJM theory.

In SYM theory, inspired by correlator/amplitude duality, the squared amplitude can be packaged into a permutation-invariant object $F_N$, represented as a sum over planar $f$-graphs. Each 
$f$-graph is a rational function of dual coordinates and enjoys an $S_N$ permutation symmetry acting on both external and internal dual points. Remarkably, by taking $n$-gon lightlike limits, $F_N$ simultaneously encodes all $n$-point $L$-loop amplitudes with 
$n+L=N$. In this framework, individual loop integrands can be efficiently extracted from 
$f$-graphs. At $n=4$, the $L$-loop integrand is obtained by identifying all four-faces of $(4+L)$-vertex
$f$-graphs, allowing the integrand to be determined in principle up to $L=12$~\cite{Bourjaily:2016evz,He:2024cej,Bourjaily:2025iad}. At five-point, the $L$-loop parity-even integrand can be extracted from five-faces of $(5+L)$-vertex $f$-graphs, while the parity-odd contribution of $L$-loop can be obtained from penta-wheel subtopologies of $(6+L)$-vertex $f$-graphs~\cite{Ambrosio:2013pba}. At higher multiplicity, although a simple graphical prescription is no longer available, it is conjectured that all-loop integrands can still be obtained from $f$-graphs by constructing suitable integrand ansatz~\cite{Heslop:2018zut}. As a result, the planar $f$-graph representation has become a powerful tool for organizing and computing loop integrands in SYM.

In contrast, loop integrands in three-dimensional ABJM theory are much less explored. Using generalized unitarity~\cite{Britto:2004nc} and positive geometry, planar integrands up to eight points at two-loop level were constructed~\cite{Bianchi:2012cq,Brandhuber:2012un,Caron-Huot:2012sos,He:2022lfz,He:2023rou}, as well as the four-point integrand up to five loops in a bipartite representation based on the ABJM amplituhedron~\cite{He:2022cup}. Meanwhile, although a fully permutation-invariant squared amplitude was also defined, several new challenges arise when attempting to extract individual amplitudes from this squared object. First, ABJM squared amplitudes exist only for even multiplicity and even loop-level, making it a priori difficult to extract the odd-loop integrand. Second, the available 
$f$-graph data are currently limited to ten points, where certain free parameters still remain due to Gram determinant identities in three dimensions.

The central question addressed in this paper is whether the three-dimensional object $F_N$ nevertheless contains complete information about individual amplitudes, or equivalently, whether one can systematically extract all scattering amplitudes directly from the squared amplitude. We show that, based on the collection of Yangian invariants and an appropriate basis of planar dual conformal integrands, it is indeed possible to disentangle the contributions of individual amplitudes from the permutation-invariant combination. Our results provide strong evidence that scattering amplitudes of arbitrary multiplicity and loop order can be extracted from three-dimensional $f$-graphs. Before moving on to our main results, let us first review the generating function for squared amplitudes in ABJM theory. 

\paragraph{Squared amplitudes in ABJM}
In chiral superspace~\cite{Elvang:2013cua}, the $n$-point superamplitude $\mathcal A_n$ of ABJM theory satisfies super-momentum conservation:
\begin{gather}
    \mathcal A_n(\lambda,\eta)=\delta^3(P)\delta^{6}(Q)A_n(\lambda,\eta),\\
    P^{\alpha\dot\alpha}=\sum_{i=1}^n(-1)^i\lambda_i^\alpha\lambda_i^{\dot\alpha},\quad Q^{\alpha A}=\sum_{i=1}^n(-1)^i\lambda_i^\alpha\eta_i^A.
\end{gather}
The Grassmann variables $\eta_i^A$ ($A=1,2,3$) encode the external states. In ABJM theory, the only non-vanishing amplitudes are those with an even number of particles, $n=2k$, and in the middle sector $A_{n=2k}\sim\eta^{3k}$. Perturbatively, \begin{equation}
 A_n=\sum_{L=0}^\infty a^L\int{\rm d}^{3L}x\,A_n^{(L)}(x_1,x_2,\ldots x_n,x_{n+1},\ldots x_{n+L}),
\end{equation}
where $a$ denotes the coupling constant; the external dual points are defined through $ p_i= x_{i+1}- x_i\equiv x_{i,i+1}$ for $i=1,\cdots, n$ which form a light-like $n$-gon, and in addition there are $L$ generic loop/internal points for $i=n{+}1, \cdots, n{+}L$; by definition, $A_n^{(L)}$ is permutation symmetric among these internal points $x_{n+1},x_{n+2},\ldots,x_{n+L}$.

The (integrand of the) squared ABJM amplitude was recently introduced in~\cite{He:2025zen} in a manner analogous to the SYM case~\cite{Heslop:2018zut,Dian:2021idl,He:2024hbb,He:2025zbz}.
It is defined as the sum over the squares of all component amplitudes, namely
\begin{equation}\label{eq:sqdef}
M_n^{(L)}:=\frac12\left.\sum_{\ell=0}^L\overline{A_n^{(\ell)}}(\lambda,\partial_\eta)A_n^{(L-\ell)}(\lambda,\eta)\right|_{\eta=0},
\end{equation}
where the conjugated amplitude $\overline{A_{n}^{(\ell)}}$ means
\begin{equation}
        \overline{A_n^{(\ell)}}(\lambda,\partial_\eta)=(-1)^\ell A_n^{(\ell)}(\lambda,\eta)\Big|_{\eta\mapsto\partial_\eta},
\end{equation}
Notice that the integrand picks up an additional $(-1)^\ell$ factor due to the charge conjugation. More importantly, this sign forces all odd-loop squared amplitudes to vanish, which plays a crucial role in the unification of loops and legs after squaring. More explicitly at the loop-level,
\begin{equation}
    A_n^{(\ell)}A_{n}^{(L-\ell)}=\sum_{\sigma\in S_L}A_n^{(\ell)}(x_{\sigma(1)},\cdots,x_{\sigma(\ell)})
    \times A_n^{(L-\ell)}(x_{\sigma(\ell+1)},\cdots,x_{(L)})
\end{equation}

For example, with the normalization $A_{4}^{(0)}=\frac{1}{\ab{12}\ab{23}}$, the four-point squared amplitudes are
\begin{equation}
    M_4^{(0)}=\frac{1}{2}\frac{1}{x_{13}^2x_{24}^2},M_4^{(2)}=A_4^{(0)}A_{4}^{(2)}-\frac{1}{2}\left[A_4^{(1)}\right]^2,M_{4}^{(4)}=A_4^{(0)}A_{4}^{(4)}-A_4^{(1)}A_{4}^{(3)}+\frac{1}{2}\left[A_{4}^{(2)}\right]^2.
\end{equation}

A remarkable feature of the squared amplitudes/cross-sections is a hidden $S_{n+L}$ permutation symmetry which unifies these objects with different $n$ and $L$. As discovered  in~\cite{He:2025zen}, there exists a single object $F_N(x_1,{\cdots},x_N)$ which is symmetric in all $N{:=}n{+}L$ points, that packages squared amplitudes, $M_n^{(L)}$, with different $n,L$ (unifying ``loops and legs''); the latter are obtained by taking $n$-gon lightlike limits similar to SYM case~\cite{Bourjaily:2016evz}:
\begin{equation}\label{eq:symFM}
\lim_{x_{12}^2,\cdots,x_{n1}^2\to0}\sigma_nF_N\equiv {\rm Res}_{n-{\rm gon}} F_N=M_n^{(L)}(x_1,\cdots,x_N),
\end{equation}
where $\sigma_n{:=}x_{12}^2x_{23}^2{\cdots}x_{n1}^2$, and taking the $n$-gon lightlike residue,  $x_{i,i{+}1}^2=0$ for $i=1,\cdots, n$, is denoted as Res$_{n-{\rm gon}}$. The remarkable property that $F_N$ is invariant under $S_N$ permutation of $x_1,{\cdots},x_N$ means that it admits a representation in terms of {\it $f$-graphs} (with weight-$3$ for ABJM, as opposed to those weight-$4$ ones in SYM)~\cite{Eden:2011we,He:2025zen}. In ABJM theory, to manifest the vanishing of odd-particle amplitudes, its  $f$-graphs cannot contain odd cycles and therefore must be bipartite. On the other hand, dual conformal invariance in 
$D=3$ requires the integrand to have conformal weight 3 at each point, which implies that the valency of each vertex in the  $f$-graph is 3. Combining these two constraints, the 
$f$-graphs are thus weight-3 {\it bipartite graphs}.~\footnote{The lowest-point $F_4{=}\frac12\,\includegraphics[align=c,scale=0.07]{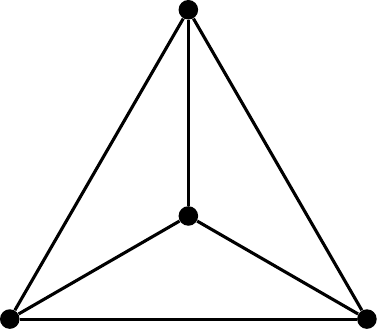}$ is an exception by being planar but not bipartite. This is similar to SYM where $F_{N\geq6}^{\rm SYM}$ are planar but $F_5^{\rm SYM}$ is nonplanar.} where each solid line $i-j$ denotes the pole $x_{ij}^2$ in the denominator and dashed line $i-j$ denotes an $x_{ij}^2$ factor in the numerator; the $S_N$ permutation invariance indicates that we must sum over all possible labeling to obtain the rational function associated with any $f$-graph. 
For example, at $N=6$, the only possible bipartite graph is the following $\mathrm{K}_{3,3}$ graph and 
\begin{equation}\label{eq:cor42}
    \begin{aligned}
    F_6&=2\times \includegraphics[align=c,scale=0.25]{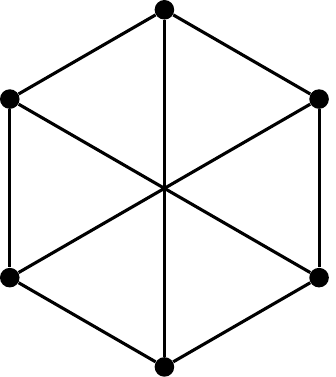}
        =2\times\left(\frac{1}{72}\frac{1}{x_{12}^2x_{23}^2x_{34}^2x_{45}^2x_{56}^2x_{61}^2x_{14}^2x_{25}^2x_{36}^2}+S_6\text{ perms}\right)\\
        &=\frac{2}{x_{12}^2x_{23}^2x_{34}^2x_{41}^2x_{13}^2x_{24}^2}\left(\frac{x_{12}^2x_{34}^2}{x_{16}^2x_{26}^2x_{35}^2x_{45}^2x_{56}^2}+\frac{x_{13}^2x_{24}^2}{x_{16}^2x_{36}^2x_{25}^2x_{45}^2x_{56}^2}+\frac{x_{14}^2x_{23}^2}{x_{16}^2x_{46}^2x_{25}^2x_{35}^2x_{56}^2}+5\leftrightarrow6\right)\\
        &+\frac{2}{x_{12}^2x_{23}^2x_{34}^2x_{41}^2x_{13}^2x_{24}^2}\left(\frac{x_{12}^2x_{23}^2x_{31}^2}{x_{16}^2x_{26}^2x_{36}^2x_{15}^2x_{25}^2x_{35}^2}+3 \text{ cyc}\right),
    \end{aligned}
\end{equation}
where the second line is the three-dimensional double triangle integral, which evaluates to elliptic polylogarithm functions as computed in~\cite{He:2023qld}, and the last line is nothing but squared triangles. After taking the lightlike limit, it gives the four-point two-loop and six-point tree-level squared amplitudes, respectively:
\begin{equation}\label{eq:f6exm}
    \begin{aligned}
        &{\rm Res}_{4-{\rm gon}}~2\times \includegraphics[align=c,scale=0.25]{images/bf6_1.pdf}{=}\frac2{x_{15}^2x_{35}^2x_{56}^2x_{26}^2x_{46}^2}{+}(5\leftrightarrow6)=M_4^{(2)}\\
        &{\rm Res}_{6-{\rm gon}}~2\times \includegraphics[align=c,scale=0.25]{images/bf6_1.pdf}{=}\frac{2}{x_{14}^2x_{25}^2x_{36}^2}=M_6^{(0)}
    \end{aligned}
\end{equation}
Meanwhile, due to the Gram identities $\det_{1\leq i,j\leq 6}x_{ij}^2=0$ in $D=3$, we can also write these bipartite $f$-graphs in terms of planar ones, for example
\begin{equation}
    2\includegraphics[align=c,scale=0.25]{images/bf6_1.pdf}=\includegraphics[align=c,scale=0.25]{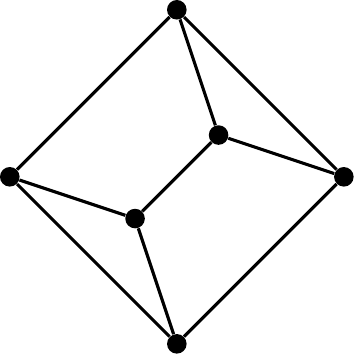}-\includegraphics[align=c,scale=0.25]{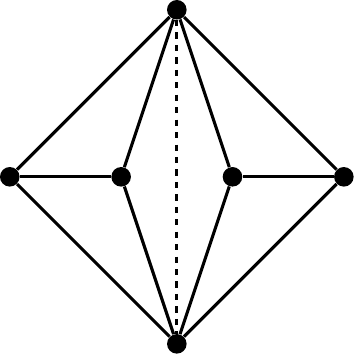}+\frac12\includegraphics[align=c,scale=0.25]{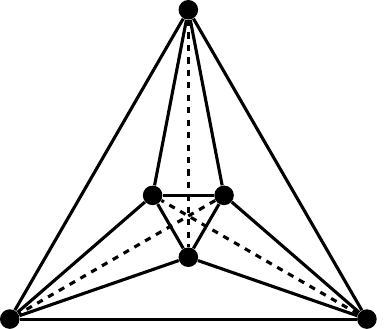},
\end{equation}
Although the vanishing of odd-particle amplitude/cut is not manifest by this planar representation, it is useful to extract four-point lower loop integrands as we show later.

This paper is organized as follows. In section \ref{sec:2}, we first explain the challenges in extracting amplitude from the squared object. Then we show how to adapt the method in~\cite{Ambrosio:2013pba} to extract four-point $(L-1)$-loop integrand from planar $f$-graph representation of $F_{4+L}$. We also comment on the relationship between squaring and taking the logarithm at four-point. In section \ref{sec:3}, we move on to the six-point extraction. After introducing a more compact formula for the squared tree amplitude, we demonstrate that the six-point amplitude up to $L=2$ can be obtained from $f$-graphs by writing down a suitable DCI integrand ansatz that satisfies cyclicity, reflection symmetry, and a soft-cut condition relating higher- and lower-loop orders. In section \ref{sec:4}, we show that the same method can be applied to the eight-point amplitude at tree-level.

\section{Four-point ABJM amplitudes from $f$-graphs}\label{sec:2}
In this section, we discuss how to extract $L$-loop integrand of four-point amplitudes from $f$-graphs with $4{+}L$ points. In $D=3$, each dual point $x_i$ can be embedded into a five-dimensional space as $X_i=(\frac12 x_i^2,1, x^\mu_i)$. The four-point integrand in ABJM theory is slightly more intricate than in SYM: in addition to its dependence on $x_{ij}^2$, it contains a parity-odd structure naturally defined in the embedding space, which is constructed from the contraction of the five-dimensional epsilon tensor with five $X_i$'s. At four-point the $\epsilon$ tensor is introduced as 
\begin{equation}
\epsilon_\ell\equiv\epsilon(\ell,1,2,3,4)=2\epsilon_{abcde}X_1^aX_2^bX_3^cX_4^dX_\ell^e,
\end{equation}
where $\ell=5,6,\cdots, 4{+}L$, and $\epsilon_\ell$ is normalized such that
\begin{equation}
    \epsilon_{\ell}^2=-x_{13}^2x_{24}^2\left(x_{24}^2x_{\ell1}^2x_{\ell3}^2+x_{13}^2x_{\ell2}^2x_{\ell4}^2\right).
\end{equation}
More generally,
\begin{equation}\label{eq:eedet}
\epsilon(a_1,a_2,a_3,a_4,*)\epsilon(b_1,b_2,b_3,b_4,*)=-\frac{1}{2}\det\left[(a_i\cdot b_j)\right].
\end{equation}
Recall that, for odd/even $L$, the $L$-loop integrand is parity odd/even, which then must contain odd/even powers of such $\epsilon$ tensors. For example, the one-loop integrand is the box integral with an $\epsilon$ numerator~\cite{Chen:2011vv}
\begin{equation}\label{eq:3dL1}
A_{4}^{(1)}=A_4^{(0)}\frac{\epsilon_\ell}{x_{\ell1}^2x_{\ell2}^2x_{\ell3}^2x_{\ell4}^2}
\end{equation}
At two-loop, the integrand can be written as a linear combination of the double box (with two such $\epsilon$ tensors contracted in the numerator) and the double triangle~\cite{Chen:2011vv}
\begin{align}\label{eq:twoloopeps}
A_4^{(2)}&=\frac{\epsilon_5\epsilon_6}{x_{15}^2x_{16}^2x_{25}^2x_{35}^2x_{36}^2x_{46}^2}-\frac{x_{13}^4}{x_{15}^2x_{16}^2x_{35}^2x_{36}^2x_{56}^2}+(5\leftrightarrow6)+(13\leftrightarrow 24)
\end{align}
\subsection{How to disentangle products of lower loops from planar $f$-graphs?}
Recall that in SYM theory, extracting the amplitude $A_4^{(L)}$ from $F_{4+L}$ is purely graphical: one simply identifies all 4-faces of planar $f$-graphs and then takes the lightlike limit~\cite{Eden:2012tu,Bourjaily:2016evz}. Moreover, the product of lower loop amplitude $A_4^{(\ell)}A_4^{(L-\ell)}$ correspond to $4$-cycles that divide the planar $f$-graph into ``interior'' and ``exterior'', containing $\ell$ and $L-\ell$ vertices, respectively.

Comparing with \eqref{eq:twoloopeps}, the first subtlety in ABJM theory is that $F_6$ is manifestly parity-even and contains no $\epsilon$-tensor. In fact, using~\eqref{eq:eedet}, one finds
\begin{equation}\label{eq:ee3d}
-2\epsilon_i\epsilon_j=x_{13}^2x_{24}^2x_{ij}^2-\text{N}^c[i{|}j],
\end{equation}
where $\text{N}^c[i{|}j]=x_{24}^2x_{i1}^2x_{j3}^2+\text{cyc}(1234)$. Substituting this into \eqref{eq:twoloopeps} yields
\begin{align}\label{eq:twolooppla}
   A_4^{(2)}&=\textcolor{red}{-\frac {x_ {2, 4}^2 x_ {1, 3}^4} {2 x_ {1, 5}^2 x_ {1, 6}^2 x_ {2, 
      5}^2 x_ {3, 5}^2 x_ {3, 6}^2 x_ {4, 6}^2}}+ \frac {x_ {2, 4}^2 x_ {1, 3}^2} {2 x_ {1, 
      6}^2 x_ {2, 5}^2 x_ {3, 5}^2 x_ {4, 6}^2 x_ {5, 
      6}^2}  + \frac {x_ {2, 6}^2 x_ {4, 5}^2 x_ {1, 3}^4} {2 x_ {1, 
      5}^2 x_ {1, 6}^2 x_ {2, 5}^2 x_ {3, 5}^2 x_ {3, 6}^2 x_ {4, 
      6}^2 x_ {5, 6}^2} \\\nonumber
      & + \frac {x_ {2, 4}^2 x_ {1, 3}^2} {2 x_ {1, 5}^2 x_ {2, 
      5}^2 x_ {3, 6}^2 x_ {4, 6}^2 x_ {5, 6}^2}- \frac {x_ {1, 
      3}^4} {2 x_ {1, 5}^2 x_ {1, 6}^2 x_ {3, 5}^2 x_ {3, 6}^2 x_ {5, 
      6}^2}+(5\leftrightarrow6)+(13\leftrightarrow 24).
\end{align}
This form only depends on $x_{i,j}^2$, though at the cost of introducing additional topologies such as the kissing-triangle (colored red in \eqref{eq:twolooppla}), which notably lacks the mutual pole $x_{56}^2$ in the denominator. The appearance of such kissing topologies is the first indication that one cannot obtain the 4-point $L$-loop integrand simply by locating 4-faces of planar $(4+L)$-vertex $f$-graphs. The topology of the graph alone does not reveal whether a given contribution corresponds to an $L$-loop integrand or to a product of lower-loop amplitudes. A similar obstruction was observed in extracting higher-multiplicity amplitudes in SYM theory~\cite{Heslop:2018zut}.

On the other hand, extracting products of lower-loop integrands cannot proceed as in four dimensions. For instance, consider the six-point planar $f$-graphs
  \begin{equation}\label{eq:ABJMF6}
F_6=\includegraphics[scale=0.2,align=c]{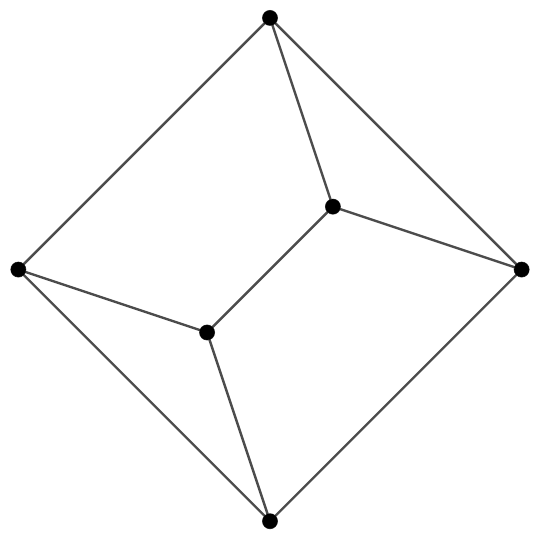}-\includegraphics[scale=0.2,align=c]{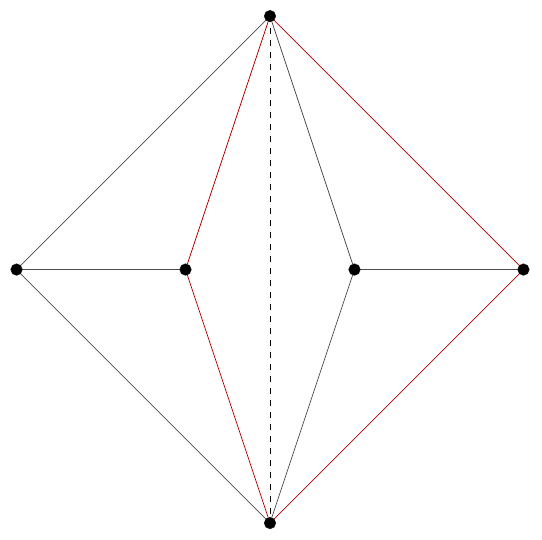}+\frac{1}{2}\includegraphics[scale=0.2,align=c]{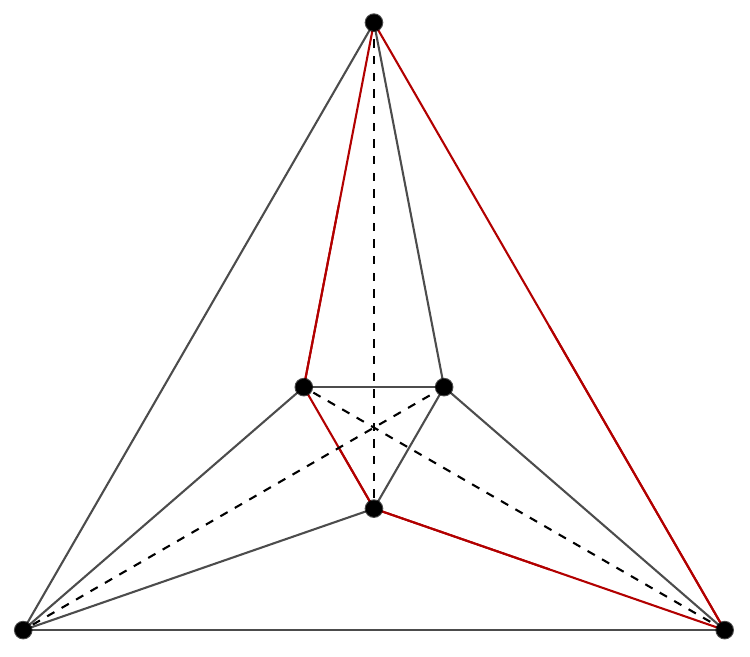},
  \end{equation}
where the red lines denote a 4-cycle that separates the planar graph into an `interior' and an `exterior', each containing exactly one vertex. By analogy with the SYM case, one might expect that this would correspond to $\left[A_{4}^{(1)}\right]^2$.  However, it actually gives
\begin{equation}
\begin{aligned}
    &-\frac{x_{13}^2x_{24}}{x_{15}^2x_{16}^2x_{26}^2x_{35}^2x_{36}^2x_{45}^2}+(5\leftrightarrow6)+(13\leftrightarrow24)+\frac{1}{2}\frac{x_{13}^4x_{24}^4x_{56}^2}{x_{15}^2x_{16}^2x_{25}^2x_{26}^2x_{35}^2x_{36}^2x_{45}^2x_{46}^2}\\
=&-2\frac{\epsilon_1\epsilon_2}{x_{15}^2x_{25}^2x_{35}^2x_{45}^2x_{16}^2x_{26}^2x_{36}^2x_{46}^2}-\frac{1}{2}\frac{x_{13}^4x_{24}^4x_{56}^2}{x_{15}^2x_{16}^2x_{25}^2x_{26}^2x_{35}^2x_{36}^2x_{45}^2x_{46}^2}\neq \left[A_{4}^{(1)}\right]^2.
    \end{aligned}
\end{equation}
where in the second line we used the identity~\eqref{eq:ee3d}.



It turns out that we can extract the $(L{-}1)$-loop integrand, which is an odd-loop integrand for $L$ even, from the planar $(4+L)$-point $f$-graphs in close analogy with the five-point parity-odd part in the SYM case. As we have checked to $L=6$ at least, one can assume that odd-loop integrands always take the form
written as 
\begin{equation}
 A_{4}^{(L-1)}=\epsilon_\ell \hat{A}_{4}^{(L-1)},  
\end{equation}
and there will never be an epsilon tensor involving two or more internal points. To isolate one-loop contribution~\eqref{eq:3dL1} from $-A_{4}^{(1)}A_{4}^{(L-1)}$ in \eqref{eq:sqdef}, by using~\eqref{eq:ee3d}, we obtain
\begin{equation}
\begin{aligned}
    -A_4^{(1)}A_4^{(L-1)}=-\frac{\epsilon_L}{x_{L1}^2x_{L2}^2x_{L3}^2x_{L4}^2}\epsilon_l \hat{A}_4^{(L-1)}
    =\frac{1}{2}\frac{x_{13}^2x_{24}^2\textcolor{red}{x_{lL}^2}+\ldots}{\textcolor{red}{x_{L1}^2x_{L2}^2x_{L3}^2x_{L4}^2}}\hat{A}_4^{(L-1)}.
\end{aligned}
\end{equation}
Therefore, to isolate the contribution of the odd-loop integrand, it is sufficient to identify all
\textbf{box-wheel} subgraphs, namely planar four-cycles that enclose one internal
vertex, with four propagators (spokes) connecting this vertex to the four corners of the box
(see Fig.~\ref{fig:boxwheel}). Each such box-wheel necessarily has a numerator line from its central vertex to some other vertex in the remainder of the $f$-graph. Removing the box-wheel and marking the endpoint of that numerator produces a new graph that contributes directly to the $(L-1)$-loop integrand which is purely parity-odd, with its coefficient doubled and the marked vertex $\ell$ carrying numerator $\epsilon_\ell$:
\begin{figure}[H]
    \centering
\includegraphics[scale=0.25]{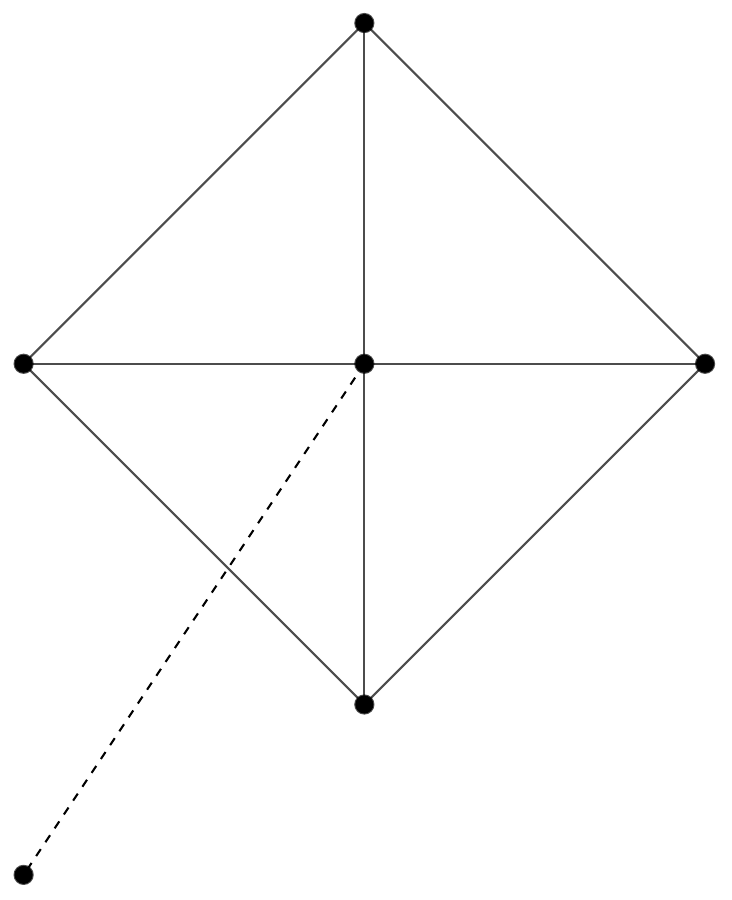}
    \caption{boxwheel}
\label{fig:boxwheel}
\end{figure}
As an example, in \eqref{eq:ABJMF6} only the last 6-point $f$-graph contains the box-wheel subgraph,
\begin{equation}
\includegraphics[align=c,scale=0.2]{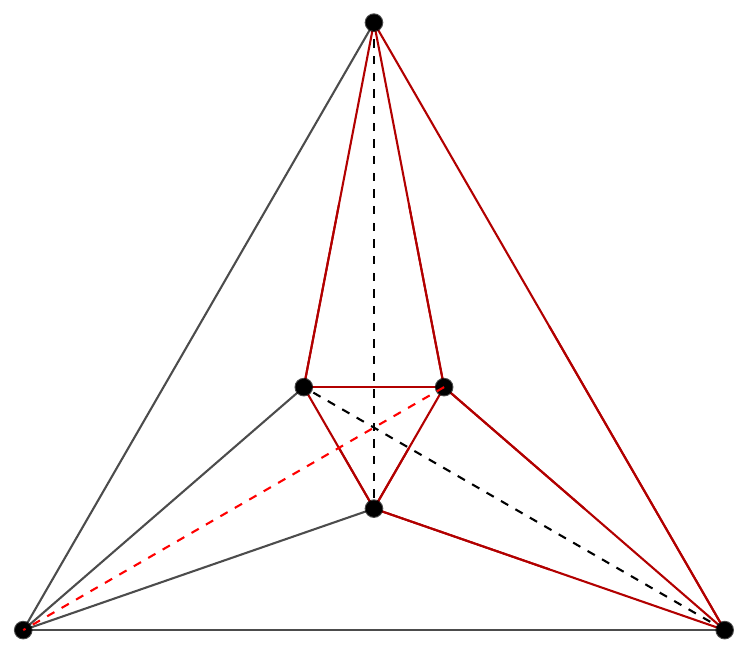}\rightarrow\includegraphics[align=c,scale=0.2]{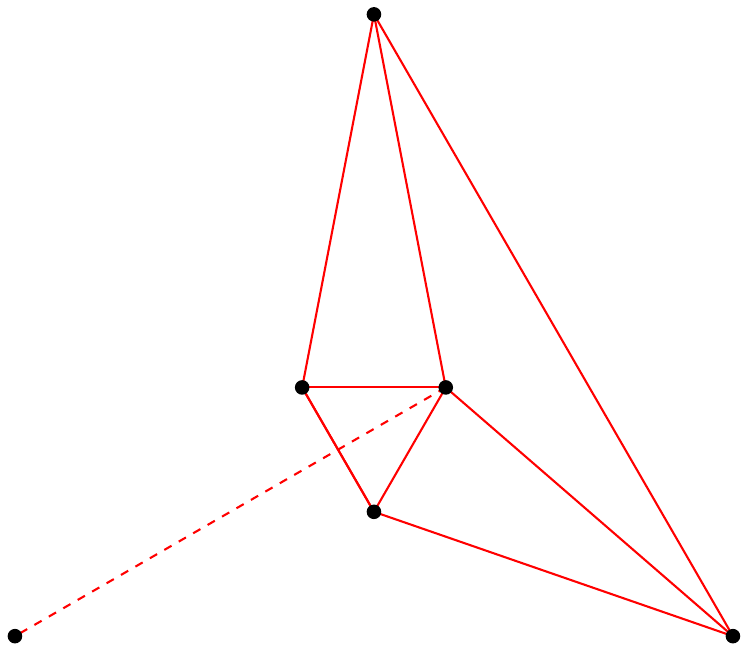}\times\includegraphics[align=c,scale=0.2]{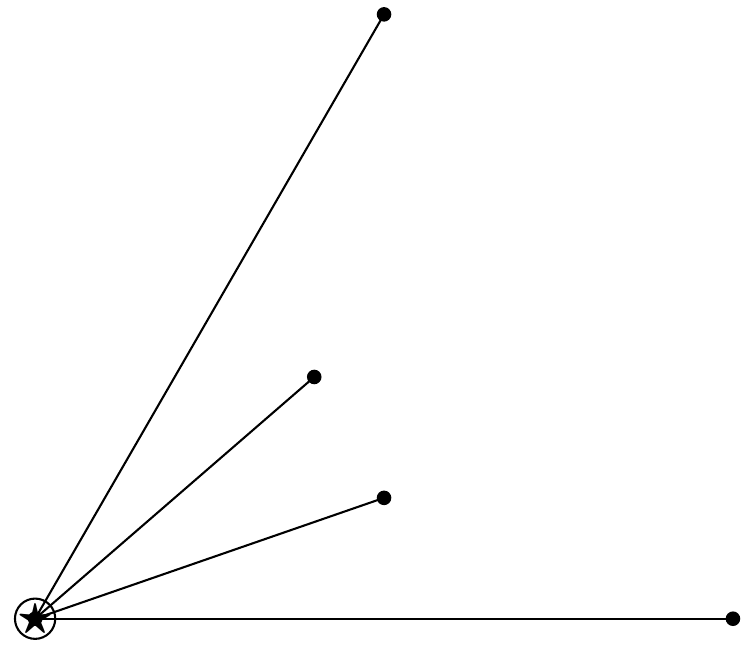},
\end{equation}
which corresponds precisely to $A_{4}^{(1)}\times A_4^{(1)}$. Similarly, at three loops there are 11 planar $f$-graphs with non-vanishing coefficients containing the box-wheel subgraph. We obtained 13 planar integrand seeds, including one kissing topology, eight ladder-type topologies, and 4 tennis-court topologies (listed in the Appendix~\ref{app:definition}). We also checked that they reproduce the result in~\cite{Bianchi:2014iia,He:2022cup}. At five loops, this procedure leads to a new planar representation of the integrand (recorded in an ancillary Mathematica file), in contrast to the bipartite and non-planar representation obtained in~\cite{He:2022cup}. Although expressed in different forms, the two representations can be explicitly checked to be equivalent.

Finally, before moving to the next subsection---where we focus on the logarithm of the amplitude, we emphasize that we can already recursively obtain the $L$-loop planar integrand from $(4+L)$-point $f$-graphs. Assuming that all lower-loop integrands are known from $f$-graphs with $N\leq 4+L$ points, then from $(4+L+2)$-point $f$-graphs, once the $(L+1)$-loop integrand is identified, one can subtract all lower-loop products from $M_{4}^{(L+2)}$ to obtain the $(L+2)$-loop amplitude with all the products of $\epsilon_i$ in odd-loop, which can be further simplified using \eqref{eq:ee3d}. We note that after doing so systematically, one obtains an even nicer representation of the planar loop integrand, which we leave to future work.

\subsection{From bipartite $f$-graphs to log of amplitudes and negative geometries}

In this subsection, we start from the bipartite $f$-graph representation of the squared amplitude in ABJM theory and connect it to the logarithm of the 4-point amplitude:
\begin{equation}
    \log R_4=\sum_{L=0}^\infty a^L(\log R_4)^{(L)},
\end{equation}
where $R_4\equiv \frac{A_4}{A_4^{(0)}}$.
Starting from a rewriting of the squared amplitudes
\begin{equation}
M_4{=}\frac{1}{2}(A_4^{(0)})^2\exp\left[\log\frac{A_4(-a)A_4(a)}{\left[A_4^{(0)}\right]^2}\right]=M_4^{(0)}\exp\left[\log{R_4(a)}+\log{R_4(-a)}\right],    
\end{equation}
we can express $M_4^{(L)}$ in terms of $(\log R_4)^{(\ell)}$ with even $\ell$ only. For example,
\begin{equation}
\begin{aligned}\label{eq:M4logexm}
    &\frac{M_4^{(0)}}{(A_4^{(0)})^2}=\frac{1}{2},\frac{M_4^{(2)}}{(A_4^{(0)})^2}=(\log R_4)^{(2)},\frac{M_4^{(4)}}{(A_4^{(0)})^2}=(\log R_4)^{(4)}+[(\log R_4)^{(2)}]^2,\\
    &\frac{M_4^{(6)}}{(A_4^{(0)})^2}=(\log R_4)^{(6)}+2(\log R_4)^{(4)}(\log R_4)^{(2)}+\frac{2}{3}[(\log R_4)^{(2)}]^3.
    \end{aligned}
\end{equation}
An important feature of $(\log{R_4})^{(\ell)}$ with $\ell>1$ is its ``bipartite pole structures", as studied in \cite{He:2022cup} using negative geometries. For example, at two loops
\begin{align}
(\log{R_{4})^{(2)}}&=\frac{M_4^{(2)}}{(A_4^{(0)})^2}=\lim_{x_{i,i+1}^2\to0}\xi_4x_{13}^2x_{24}^2\times2\includegraphics[align=c,scale=0.2]{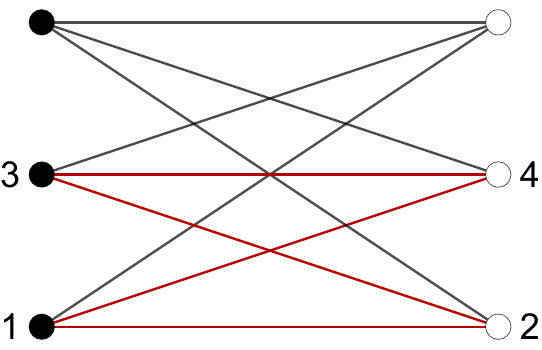}\\\nonumber
&=\includegraphics[align=c,scale=0.3]{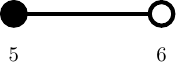}{+}\includegraphics[align=c,scale=0.3]{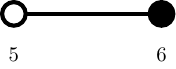}=\frac{2x_{13}^2x_{24}^2}{x_{51}^2x_{53}^2x_{62}^2x_{64}^2x_{56}^2}\left(5\leftrightarrow6\right),
\end{align}
where each black node $i$ represents the pole $s_i:=x_{i1}^2x_{i3}^2$, while each white node $j$ represents $t_j=x_{j2}^2x_{j4}^2$; the link between two internal dual points denotes the pole $x_{ij}^2$. For higher loops, there will be more than one bipartite topologies, and these bipartite graphs from negative geometries fully determine the pole structures of $\log R_4$.

Using \eqref{eq:M4logexm}, we can obtain the log of amplitudes recursively at even loops from the squared amplitudes. Inspired by \cite{He:2022cup}, we can regroup the full results into connected bipartite topologies. To achieve this, we can write down an ansatz for each bipartite topology which has the correct DCI weight and symmetry properties using $\epsilon_i$ and $x_{ij}^2$ (not including the mutual ones).  For example, at four loops, there are 3 bipartite topologies,{\it i.e.} chain, star and box topology.
\begin{equation}
    (\log{R_4})^{(4)}=-C-S+B,
\end{equation}
where
\begin{figure}[H]
    \centering
     \includegraphics{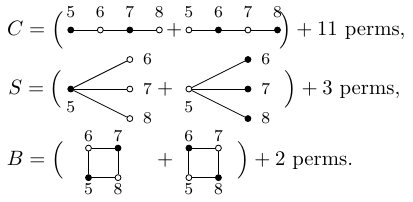}
    \label{4loops}
\end{figure}

By matching it to the result from the bipartite $f$-graph, we are left with 2 free parameters due to \eqref{eq:ee3d}. This degree of freedom does not affect the final result of $(\log{R_4})^{(4)}$ but merely about how we regroup the integrand. To fully fix the remaining parameters, we can use the box vanishing cut to fix them and reproduce the result from negative geometry in \cite{He:2022cup}. Explicitly, for any box subtopology, negative geometry tells us the following cut will vanish
\begin{equation}
x_{ij}^2=x_{jk}^2=x_{j2}^2=x_{j4}^2=x_{i1}^2=x_{k3}^2=0.
\end{equation}

Similarly, at six-loop, we have 17 bipartite topologies, including 6 tree topologies whose canonical form is known \cite{He:2022cup} and 11 loop topologies. There are still free parameters after matching the result from the $f$-graph, and the number of free parameters can be reduced using geometry cuts including the box vanishing cut. For example, the result of the hexagon topology can be fixed
\begin{equation}
    \begin{aligned}
    &\includegraphics[align=c]{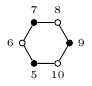}{=}2^3\frac{\left(2^3\epsilon_{5,6,\ldots,10}{-}(\epsilon_{5,7}\text{N}^t[6,9{|} 8,10]{+}
\epsilon_{6,8}\text{N}^t[7,10{|}9,5]+\text{cyclic by }2){-}\text{N}[5,7,9{|}6,8,10]\right)}{s_5t_6s_7t_{8}s_{9}t_{10}D_{5,6}D_{6,7}D_{7,8}D_{8,9}D_{9,10}D_{10,5}},
    \end{aligned}
\end{equation}
where $\epsilon_A=\prod_{i\in A}\epsilon_i,\mathrm{N}^{t}[A|B]=(x_{13}^2)^{\abs{A}}\prod_{i\in A,j\in B}x_{i2}^2x_{j4}^2+(i\leftrightarrow j),\mathrm{N}[A|B]=\mathrm{N}^{t}[A|B]+(1,3\leftrightarrow2,4)$, while there are still free parameters in other topologies such as $\mathrm{K}_{24},\mathrm{K}_{33}$ and etc. We record the complete result in the Mathematica file.

Before we move on to higher-point amplitude, we make an all-loop comment on bipartite $f$-graphs. For each bipartite graph, the vertices are divided into two sets $\{I,J\}$ where $\abs{I}<\abs{J}$. To satisfy DCI weight, $\abs{I}\geq3$ and when $\abs{I}=3$, the only bipartite $f$-graph is $K_{3,L+1}$.
\begin{equation}
    \includegraphics[align=c,scale=1]{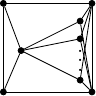}=\frac{x_{13}^{L-2}x_{51}^{L-2}x_{53}^{L-2}}{\prod_{\substack{i=1,3,5\\ j=2,4,6,\ldots,4+L}}x_{ij}^2}+\text{inequivalent perms}.
\end{equation}
This $f$-graph is one-to-one corresponding to the bipartite star negative geometry at $L$-loop, whose canonical form is
\begin{equation}
    \includegraphics[align=c,scale=1.2]{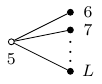}=\frac{2^{L-1}x_{24}^2x_{13}^{L}x_{51}^{L-2}x_{53}^{L-2}}{x_{52}^2x_{54}^2\prod_{j=6}^{L+4}x_{5j}^2x_{j1}^2x_{j3}^2}
\end{equation}
Therefore, the negative geometry tells us the coefficient of the bipartite $f$-graph $K_{3,L+1}$ is $2^{L-1}$. At $L=4$, it also gives the largest coefficient of bipartite $f$-graphs. It will be interesting to fix the Gram by other graphical rules at $L=6$ to see whether the largest coefficient is also $2^{5}$ given by $\mathrm{K}_{3,11}$.
\section{Six-point ABJM amplitudes from $f$-graphs}\label{sec:3}
\subsection{Squaring tree amplitudes and leading singularities}
In ABJM theory, the $(n=2k)$-point tree-level amplitude is associated with $(n-3)$-dimensional cells of the orthogonal Grassmannian $OG_+(k,2k)$. More precisely, the BCFW building blocks appearing in the tree-level recursion can be written as residues of the integral~\cite{Lee:2010du,Huang:2014xza}
\begin{equation}
    \mathcal A_{n=2k}=\int\frac{{\rm d}^{k\times2k}C}{GL(k)}\frac{\delta^{\frac{k(k+1)}2}(C\cdot C^T)}{\prod_{i=1}^k M_i}\delta^{2k{\mid} 3k}(C\cdot\Lambda)
\end{equation}
where the dot products are defined according to the $n$-dimensional splitting signature metric $\eta={\rm diag}(1,-1,\cdots,+1,-1,+1)$, the kinematic data including the Grassmann variables is encoded in $\Lambda_i=(\lambda_i,\eta_i)$, and the minors are defined as $M_i=(i, i{+}1, \cdots , i{+}k{-}1)$. Localizing $M_i=0$ restricts it to an $(n-3)$-dimensional subspace by evaluating the residues on the loci of $M_i=0$, corresponding to cells with co-dimension $\frac{(k-2)(k-3)}{2}$ of the positive orthogonal Grassmannian $\mathrm{OG}_+(k,2k)$. For example, at $n=4,6$ the tree amplitude is given by the top cell, while for $n=8$ the BCFW terms are given by co-dimension one cells corresponding to $M_i=0$ for $i=1,2,3,4$. Also note that since the orthogonal constraint is quadratic in
$C$, $2\times 2^{k-2}$ solutions are split into positive and negative branches, which satisfy $\frac{M_I}{M_{\bar{I}}}=\pm 1$.

From (super)momentum conservation, the orthogonal Grassmannian matrix $C$ can be fixed to be
\begin{equation}
    C=\left(
    \begin{aligned}
        &\lambda\\
        &\hat{C}
    \end{aligned}
    \right),
\end{equation}
then we write the full amplitudes as sum over cells $\sigma$, the branches $\pm$ and solutions:
\begin{equation}
\begin{aligned}
\mathcal{A}_n=A_n\delta^3(P)\delta^6(Q)&\equiv \sum_{\sigma}\sum_{\pm}\sum_{\text{sol}}f_{\sigma,\pm,\text{sol}}\delta^{k-2}(\hat{C}_{\sigma,\pm,\text{sol}}\cdot\eta)\delta^3(P)\delta^6(Q)\\\nonumber
&\equiv\delta^3(P)\delta^6(Q)\sum_{\sigma}\sum_{\pm}\sum_{\text{sol}}\mathrm{LS}_{\pm,\text{sol}}[\sigma_i]
\end{aligned}
\end{equation}

When computing the product of two super-functions according to \eqref{eq:sqdef}, we have
\begin{equation}\label{eq:sqdet}
\begin{aligned}
    f\delta^{k-2}(\hat{C}\cdot\eta)\star f^\prime\delta^{k-2}(\hat{C}^\prime\cdot\eta)
&=ff^\prime\prod_{I=1}^3\prod_{j=1}^{k-2}\sum_{\alpha=1}^ni^{\text{odd}(\alpha)}\hat{C}^\prime_{j,\alpha}\partial_{\eta_{\alpha,I}}\prod_{I^\prime =1}^3\prod_{j^\prime =1}^{k-2}\sum_{\alpha^\prime =1}^ni^{\text{odd}(\alpha^\prime)}\hat{C}_{j,\alpha}\eta_{\alpha^\prime,I^\prime}\\
&=ff^\prime\prod_{I=1}^3\sum_{\alpha\in S_{k-2}}\mathrm{sgn}(\sigma)\prod_{j=1}^{k-2}\sum_{\alpha=1}^n (-1)^\alpha \hat{C}_{j\alpha} \hat{C}^\prime_{\sigma(j),\alpha}\\
&=ff^\prime \det(\hat{C}\cdot \hat{C}^{\prime \mathrm{T}})^3,
\end{aligned}
\end{equation}
where in the second line we used $\partial_{\eta_{\alpha,I}}\eta_{\alpha^\prime,I^\prime}=\delta_{\alpha,\alpha^\prime}\delta_{I,I^\prime}$ and the anti-community property of Grassmannian variables. An immediate consequence is that the square of any super-function itself vanishes due to the orthogonal condition $\hat{C}\cdot \hat{C}^T=0$. 

At six points, only the top cell contributes which can be parametrized as
\begin{equation}
C_+=   
\begin{pmatrix}
&\lambda_1&\lambda_2&\lambda_3&\lambda_4&\lambda_5&\lambda_6\\
&\langle35\rangle&\langle24\rangle&\langle51\rangle&\langle62\rangle&\langle13\rangle&\langle46\rangle
\end{pmatrix},\quad
C_-=   
\begin{pmatrix}
&\lambda_1&\lambda_2&\lambda_3&\lambda_4&\lambda_5&\lambda_6\\
&\langle35\rangle&-\langle24\rangle&\langle51\rangle&-\langle62\rangle&\langle13\rangle&-\langle46\rangle
\end{pmatrix},
\end{equation}
where $\pm$ corresponds to the positive/negative branch, and they give two leading singularities
\begin{equation}
    \mathrm{LS}_{\pm}=\frac{\delta^3(C_{\pm}\cdot\eta)}{2c_{1}^{\pm}c_2^{\pm}c_3^{\pm}},
\end{equation}
where
\begin{equation}
    \begin{aligned}
    &c_{1}^{\pm}=-\langle5{\mid}135{\mid}2\rangle\mp\langle13\rangle\langle46\rangle\\
&c_{2}^{\pm}=\pm\langle6{\mid}246{\mid}3\rangle\mp\langle24\rangle\langle51\rangle\\
    &c_{3}^{\pm}=-\langle1{\mid}135{\mid}4\rangle\mp\langle35\rangle\langle62\rangle\\
    \end{aligned}
\end{equation}
After stripping of the $\delta^3(P)\delta^6(Q)$, the tree amplitude $A_6$ is proportional the sum of two branches
\begin{equation}\label{eq:A6a}
A_6=\alpha(\mathrm{LS}_++\mathrm{LS}_-)
\end{equation}
The six-point case provides a trivial example of extracting tree amplitude similar to \cite{Heslop:2018zut}, where we only need to fix the overall coefficient $\alpha$ by comparing it to $M_6^{(0)}$~\eqref{eq:f6exm}. Using~\eqref{eq:sqdet}, the square of $\mathrm{LS}_{\pm}^2$ vanishes, and only the product between positive and negative branches contributes to the squared amplitudes
\begin{equation}
M_6^{(0)}=\frac{1}{2}A_6^2=\alpha^2\mathrm{LS}_+\mathrm{LS}_-=\frac{\alpha^2\det(C_+\cdot C_-)^3}{4(c_{1}^+c_{1}^-)(c_{2}^+c_{2}^-)(c_{3}^+c_{3}^-)}.
\end{equation}
The determinant in the numerator becomes
\begin{equation}
    -\langle13\rangle^2-\langle35\rangle^2-\langle51\rangle^2-\langle24\rangle^2-\langle46\rangle^2-\langle62\rangle^2=-s_{135}
\end{equation}
while the factors in the denominators can be simplified to be
\begin{equation}
        c_{1}^+c_{1}^-=x_{14}^2s_{135},c_{2}^+c_{2}^-=-x_{25}^2s_{135},c_{3}^+c_{3}^-=x_{36}^2s_{135}.
\end{equation}
Therefore the square of ansatz \eqref{eq:A6a} becomes
\begin{equation}
    M_6^{(0)}=\frac{2\alpha^2}{x_{14}^2x_{25}^2x_{36}^2}.
\end{equation}
Comparing to \eqref{eq:f6exm}, we can obtain $\alpha=\pm 1$, which gives the correct result \cite{Huang:2012hr} by choosing the positive sign. The correlator can
never predict this sign since the duality \eqref{eq:symFM} only gives the square
of the amplitude. If we had chosen the wrong sign here, it would affect higher-loop amplitudes. 

For higher loops, it is convenient to introduce the shifted tree amplitude $A_6^{\text{s}}(\bar12\bar34\bar56)=A_6(\bar23\bar45\bar61)=-i(\mathrm{LS}_+-\mathrm{LS}_-)$, whose square also gives $M_6^{(0)}$. Meanwhile the mixed product $A_6\star A_6^{\text{s}}=0$. This vanishing result is natural: the product must yield a parity-even object, whereas $A_6$ and $A_6^{\text{s}}$ have opposite parity.

\subsection{Extracting one and two-loop integrands}
Following the strategy of \cite{Heslop:2018zut}, we attempt to extract the six-point loop integrand from the $f$-graph data by introducing the ansatz
\begin{equation}\label{eq:ans6l}
    A_6^{(L)}=A_6 I^{(L)}+A_6^{\text{s}} I_{\text{s}}^{(L)}.
\end{equation}
In addition to having the correct dual conformal weight, the ansatz must obey several physical constraints such as cyclicity, reflection and parity\cite{Bargheer:2012cp,He:2022lfz}
\begin{equation}\label{eq:sym}
\begin{aligned}
    &A_n(\bar{1}2\bar{3}\ldots n)=(-1)^{\frac{n}{2}-1}A_n(\bar{3}4\bar{5}\ldots2)\\
    &A_n(\bar{1}2\bar{3}\ldots n)=(-1)^{\frac{n(n-2)}{8}+L}A_n(\bar{1}n\ldots 4\bar{3}2)\\
    &A_n(\ldots,-\Lambda_i,\ldots)=(-1)^{F_i}A_n(\ldots,\Lambda_i,\ldots)
\end{aligned},
\end{equation}
where $F_i$ denotes fermion number of leg $i$. These relations implies $I^{(L)}$ and $I_{\text{s}}^{(L)}$ are invariant under cyclic by 2 and pick up an $(-1)^{L}$ under reflection. Also $I^{(L)}$ should be parity even (odd) at even (odd) loop, while $I_{\text{s}}^{(L)}$ has the opposite parity. 

Let us show that, at $L=2$, comparing the squared integrand ansatz with the result obtained from the $f$-graph completely fixes the one-loop integrand as well as the parity-even part of the two-loop integrand. More generally, when squaring the ansatz~\eqref{eq:ans6l} at $L$ loops, the even-loop squared amplitude $M_6^{(L)}$ contains contribution from $I^{(L)},I^{(L-1)},I_{\text{s}}^{(L)}$ as follows
\begin{equation}
A_6I^{(L)}*A_6I^{(0)}+A_6 I^{(L-1)}*A_6 I^{(1)}+A_6^\text{s} I_\text{s}^{(L-1)}*A_6^\text{s} I_{\text{s}}^{(1)}
\end{equation}
which gives some ambiguity $I_{\text{parity odd}}$ and $I_{\text{parity even}}$ that is invisible in $f$-graph
\begin{equation}\label{eq:ambi}
\begin{aligned}
    &I^{(L-1)}\rightarrow I^{(L-1)}+I_{\text{parity odd}},I_s^{(L-1)}\rightarrow I_s^{(L-1)}+I_{\text{parity even}}\\
    &I^{(L)}\to I^{(L)}-I_{\text{parity odd}}I^{(1)}-I_{\text{parity even}}I_s^{(1)},
    \end{aligned}
\end{equation}
where $I_{\text{parity odd}}$ and $I_{\text{parity even}}$ are any combination $(L{-}1)$-loop integrands which are parity odd/even. In fact, \eqref{eq:ambi} is the only form of ambiguity arising from the duality at the loop level.

More explicitly, at two loops \eqref{eq:sqdef} becomes
\begin{equation}
    \frac{M_{6}^{(2)}}{M_6^{(0)}}=I^{(2)}-\frac{1}{2}(I_s^{(1)})^2-\frac{1}{2}(I^{(1)})^2
\end{equation}

Then we write down a linear independent basis of loop integrands. At one loop, we only have parity-even triangles (with $(i,j,k)=(1,3,5)$ or $(2,4,6)$), and parity-odd boxes with $\epsilon$ numerator ($1\leq i<j<k<m\leq 6$)
\begin{equation}
\begin{aligned}
&\text{tri}[i,j,k]=\frac{\sqrt{x_{ij}^2x_{jk}^2x_{ki}^2}}{x_{\ell i}^2x_{\ell j}^2x_{\ell k}^2}\equiv\frac{(i\cdot j\cdot k)}{x_{\ell i}^2x_{\ell j}^2x_{\ell k}^2}\qquad \text{parity-even triangle}\\
    &\text{box}[i,j,k,m]=\frac{\epsilon(\ell,i,j,k,m)}{x_{\ell i}^2x_{\ell j}^2x_{\ell k}^2x_{\ell m}^2}\qquad \text{parity-odd box}
\end{aligned}    
\end{equation}
Among the 15 boxes, five linear relations follow from the 5-term Schouten identity of the 5D Levi-Civita tensor,
\begin{equation}
    \text{box}[i,j,k,l]-\text{box}[j,k,l,p]+\text{box}[k,l,p,i]-\text{box}[l,p,i,j]+\text{box}[p,i,j,k]=0,
\end{equation}
leaving an ansatz with $2+10$ coefficients (where $I_o$, $I_e$ denotes these triangles and boxes):
\begin{equation}
A_6^{(1)}=A_6\sum_{i=1}^{10}\mathrm{co}_i^{(1)}\mathrm{Io}_i^{(1)}+A_6^{\text{s}}\sum_{i=1}^2 \mathrm{ce}_i^{(1)}\mathrm{Ie}_i^{(1)}.
\end{equation}

At $L=2$, the numerator may contain both $x_{ij}^2$ and the contraction of two $\epsilon$ tensor $\epsilon(a,i,j,k,*)\epsilon(b,l,m,n,*)$. As pointed out in \cite{Caron-Huot:2012sos}, the most general topology can be taken to be double-box integrals. Topologies such as penta-box can be reduced to double-box integrals. Imposing the correct dual conformal weight yields $739$ planar connected DCI integrands, subject to 54 linear relations from \eqref{eq:eedet}, leaving an ansatz with $685$ coefficients:
\begin{equation}
I^{(2)}=\sum_{i=1}^{685}\mathrm{ce}_i^{(2)}\mathrm{Ie}_i^{(2)}
\end{equation}

After imposing cyclic and reflection symmetry \eqref{eq:sym}, the one-loop ansatz contains $5=2+3$ free parameters, while the two-loop ansatz contains $135$ parameters. Matching against the bipartite $f$-graph data further reduces the numbers, but still leaves a two-dimensional ambiguity. Instead of following the strategy of \cite{Heslop:2018zut}, which would require accessing $10$-point $f$-graph data and thus performing computation at $L=3,4$ in ABJM theory, we instead make use of the soft cut \cite{Huang:2012hr, He:2022lfz}, which acts as a recursion relation for the loop integrand and thus reduces the ansatz space
\begin{equation}
    \left.A_n^{(L)}\right|_{x_{ai-1}^2=x_{ai}^2=x_{ai+1}^2=0}=(-1)^i A_n^{(L-1)}.
\end{equation}
By using soft cut, we observed that the two remaining parameters are fixed. For example, the one-loop part can be written as follows
\begin{equation}
    \begin{aligned}
        I_{\text{s}}^{(1)}&=\delta_1 \text{tri}[1,3,5]+\delta_2\text{tri}[2,4,5]\\
        I^{(1)}&=2\left(\text{box}[1,3,4,5]+\text{box}[1,4,5,6]-\text{box}[1,2,3,4]-\text{box}[1,2,4,6]\right)\\
        &+\beta\left(2\text{box}[1,2,4,6]+\text{box}[1,3,5,6]+\text{box}[1,2,3,5]\right.\\
        &\quad\left.-\text{box}[1,2,3,6]-\text{box}[1,2,4,5]-\text{box}[1,2,5,6]-\text{box}[1,3,4,6]\right),
    \end{aligned}
\end{equation}
where the remained coefficients $\beta,\delta_1,\delta_2$ satisfies following quadratic equations
\begin{equation}
    \begin{cases}
        &4(\beta-1)^2=\delta_1^2+8\\
        &\delta_1^2=\delta_2^2,\delta_1\delta_2=4
    \end{cases}.
\end{equation}
This system admits 8 solutions, {\it i.e.}
\begin{equation}
    (\beta,\delta_1,\delta_2)=(0,\pm2i,\mp2i),(2,\pm2i,\mp2i),(1+\sqrt{3},\pm2,\pm2),(1-\sqrt{3},\pm2,\pm2).
\end{equation}
The first solution, $(\beta,\delta_1,\delta_2)=2i(0,1,-1)$, reproduces the correct one-loop integrand of \cite{Huang:2012hr}. Moreover, once this one-loop choice is fixed, the two-loop result automatically matches the parity-even part obtained in \cite{Huang:2012hr} and can be obtained via the Mathematica file. In this sense, the ambiguity is purely a one-loop effect and can be fixed once and for all at the one-loop level. Our results provide evidence for the following conjecture: the parity even part of $A_6^{(L)}$ {\it i.e.} $I^{(L)}$ and the entire $(L-1)$-loop data can be extracted from $M_6^{(L)}$ up to a choice of solution to a quadratic system.

We also comment on the log of the six-point amplitude. We can define
\begin{equation}
\begin{aligned}
    &\log I=\sum_{L=0}^\infty a^L(\log I)^{(L)}\\
    &\log I_{\text{s}}=\sum_{L=0}^\infty a^L(\log I_{\text{s}})^{(L)}.
\end{aligned}
\end{equation}
At two-loop, \eqref{eq:sqdef} implies $M_6^{(2)}=\frac{1}{2}(\log I)^{(2)}+\frac{1}{4}\left((\log I_\text{s})^{(1)}\right)^2$. After subtracting the $I_{\text{s}}^{(1)}$ contribution, we obtain the two-loop logarithm of the parity-even integrand,
\begin{equation}
    (\log I)^{(2)}=\sum_{i=1}^6[i,i+2{\mid} i+1,i+3]+[13{\mid}46]+[35{\mid}62]+[51{\mid}24]+\left([135{\mid}24]+5 \text{cyclic}\right)+[135{\mid}246].
\end{equation}
Here, the building blocks with bipartite pole structures are
\begin{align}
    &[ij{\mid} kl]=2\frac{x_{ij}^2x_{kl}^2-2x_{ik}^2x_{jl}^2}{x_{7i}^2x_{7j}^2x_{8k}^2x_{8l}^2x_{78}^2}+7\leftrightarrow8\\\nonumber
    &[ijk{\mid} lm]=2\frac{x_{7l}^2x_{im}^2x_{jk}^2+\text{cyclic}(i,j,k)+l\leftrightarrow m}{x_{7i}^2x_{7j}^2x_{7k}^2x_{8l}^2x_{8m}^2x_{78}^2}+7\leftrightarrow8\\
    &[135{\mid} 246]=2\frac{x_{18}^2x_{47}^2x_{25}^2x_{36}^2+x_{38}^2x_{67}^2x_{14}^2x_{25}^2+x_{58}^2x_{27}^2x_{14}^2x_{36}^2+x_{78}^2 P_6}{x_{71}^2x_{73}^2x_{75}^2x_{82}^2x_{84}^2x_{86}^2x_{78}^2}+7\leftrightarrow8,
\end{align}
where $P_6=4(1\cdot3\cdot5)(2\cdot 4\cdot 6)+x_{13}^2x_{25}^2x_{46}^2+x_{13}^2x_{25}^2x_{46}^2+x_{15}^2x_{24}^2x_{36}^2-2x_{14}^2x_{25}^2x_{36}^2$.

\section{Eight-point ABJM amplitudes from $f$-graphs}\label{sec:4}
At eight points, we focus on the extraction of the tree amplitude, which provides a more nontrivial example compared to the six-point tree-level case. We consider all codimension-1 cells defined by imposing $M_i=0$ for $i=1,2,3,4$. Among these four cells, only three are linearly independent. Apart from the positive/negative branches, each branch now contains two solutions. Our ansatz for the tree amplitude is therefore
\begin{equation}
A_8=\sum_{i=1}^3\alpha_i\sum_{\pm}\sum_{\text{sol}}\mathrm{LS}_{\pm,\text{sol}}[\sigma_i],
\end{equation}
where $\sigma_i$ denotes the cell where $M_i=0$ and the corresponding leading singularity is given in \cite{He:2022lfz}. When squaring the ansatz, in addition to the fact that $\mathrm{LS}_{\pm,\text{sol}}[\sigma_i]^2=0$, we also checked numerically that 
\begin{equation}\label{eq:vanish8}
    \mathrm{LS}_{+,\text{sol}}[\sigma_i]\star\mathrm{LS}_{-,\text{sol}^\prime}[\sigma_j]=0,
\end{equation}
where $\sigma_i$ and $\sigma_j$ denote two cells(not necessarily distinct). As follows from \eqref{eq:sqdet}, this vanishing occurs because $\hat{C}\cdot \hat{C}^{\prime \mathrm{T}}$ is not a full-rank matrix. Recall in the six-point case, only the product between the positive and negative branches gives a non-vanishing contribution to the squared amplitude. Interestingly, this vanishing pattern alternates with $k$.
We have checked this alternating behavior numerically up to ten points.
More precisely, besides the trivial vanishing
$\mathrm{LS}_{\pm,\text{sol}}[\sigma_i]^2=0$, we find the following systematic pattern:
\begin{equation}\label{eq:vanishLS}
    \begin{aligned}
        &\mathrm{LS}_{+,\text{sol}}[\sigma_i]\star\mathrm{LS}_{-,\text{sol}^\prime}[\sigma_j]=0,\quad k \text{ even}\\
        &\mathrm{LS}_{+,\text{sol}}[\sigma_i]\star\mathrm{LS}_{+,\text{sol}^\prime}[\sigma_j]=\mathrm{LS}_{-,\text{sol}}[\sigma_i]\star\mathrm{LS}_{-,\text{sol}^\prime}[\sigma_j]=0,\quad  k \text{ odd}.
    \end{aligned}
\end{equation}
We leave the general proof in the Appendix \ref{app:proofvanish}.

From \eqref{eq:vanish8}, we can see that after summing over branches and solutions, the square of any cell and the product of two cells become
\begin{equation}
\begin{aligned}
    &\frac{1}{2}\mathrm{LS}[\sigma_i]^2=\frac{1}{2}\left(\sum_{\pm}\sum_{\text{sol}}\mathrm{LS}_{\pm,\text{sol}}[\sigma_i]\right)^2=\mathrm{LS}_{+,\text{sol}_1}[\sigma_i]\star\mathrm{LS}_{+,\text{sol}_2}[\sigma_i]+(+\leftrightarrow-)\\
    &\frac{1}{2}\mathrm{LS}[\sigma_i]\star\mathrm{LS}[\sigma_j]=\sum_{\text{sol}}\mathrm{LS}_{+,\text{sol}}[\sigma_i]\star\sum_{\text{sol}^\prime}\mathrm{LS}_{+,\text{sol}^\prime}[\sigma_j]+(+\leftrightarrow-)\\
    \end{aligned}
\end{equation}
On the other hand, after taking the lightlike limit of the eight-point bipartite $f$-graphs
\begin{equation}
F_8=\includegraphics[align=c,scale=0.25]{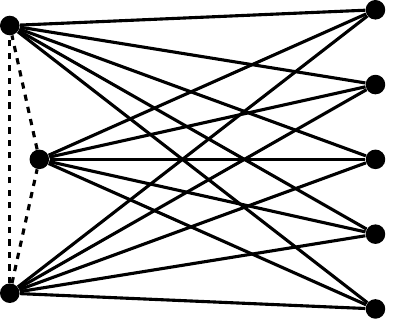}-4\includegraphics[align=c,scale=0.25]{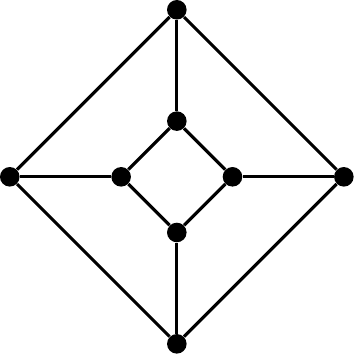}+2\includegraphics[align=c,scale=0.25]{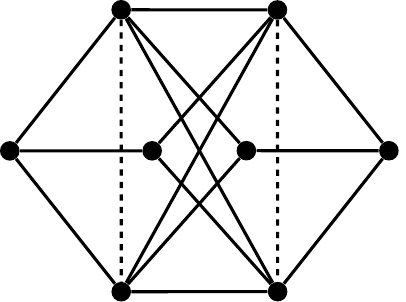},
\end{equation}
we obtain the eight-point squared tree amplitude
\begin{equation}
\begin{aligned}
    M_8^{(0)}&{=}\frac{2}{x_{14}^2x_{36}^2x_{58}^2x_{72}^2}\left({-}2{+}\frac{1}{v_2}{+}\frac{1}{v_4}{+}u_1v_1u_6{+}u_3v_3u_8+u_4u_7v_3{+}u_2u_5v_1{+}\frac{(u_1{+}u_5)v_1}{v_2}{+}\frac{(u_3{+}u_7)v_3}{v_4}\right)\\
    &{+}\frac{2}{x_{25}^2x_{47}^2x_{61}^2x_{83}^2}\left({-}2{+}\frac{1}{v_1}{+}\frac{1}{v_3}{+}u_2v_2u_7{+}u_4v_4u_1{+}u_5u_8v_4{+}u_3u_6v_2{+}\frac{(u_4{+}u_8)v_4}{v_1}{+}\frac{(u_2{+}u_6)v_2}{v_3}\right),
\end{aligned}
\end{equation}
where the cross ratios are defined~\cite{He:2022lfz} as $u_i=\frac{x_{ii+1}^2x_{i+3,i+7}^2}{x_{ii+3}^2x_{i+2,i+7}^2},v_i=\frac{x_{ii+3}^2x_{i+4,i+7}^2}{x_{ii+4}^2x_{i+3,i+7}^2}$.

By numerically comparing the squared ansatz and the squared amplitude from the $f$-graph, we obtained 
\begin{equation}
    \alpha_1=\alpha_3=\pm1,\alpha_2=0,
\end{equation}
and hence
\begin{equation}
A_8=\pm(\sum_{\pm}\sum_{\text{sol}}\mathrm{LS}_{\pm,\text{sol}}[\sigma_1]+\sum_{\pm}\sum_{\text{sol}}\mathrm{LS}_{\pm,\text{sol}}[\sigma_3]).
\end{equation}
By choosing the positive sign, we reproduce the result in~\cite{Gang:2010gy, He:2022lfz}. We note that the cyclic-by-two symmetry, although not imposed at tree level, emerges manifestly in the final result.




\section{Discussions and Outlook}
The hidden permutation symmetry of squared amplitudes provides a new way for studying ABJM amplitudes and reveals remarkable unification and simplifications: different cross-sections with fixed $n+L$ are combined in a single function, $F_N$, which is a linear combination of bipartite $f$-graphs (they can also be rewritten using planar $f$-graphs via Gram identities). In this paper, we provide some evidence for the conjecture that individual amplitudes can be extracted from these squared ones. As a proof of concept,  by constructing an ansatz via squaring leading singularities (or Yangian invariants), we show how to extract from $F_N$ with $N\leq 8$ the six-particle amplitudes at tree-level,
one-loop, and (the parity-even part of) two-loop case, as well as eight-point tree amplitudes. Moreover, the four-point case is similar to the $n=5$ case for SYM, and we extract up to $L=6$ results from $F_{10}$: not only have we obtained a new planar representation for five-loop integrands directly, but we have also been able to provide new results for the log of four-point amplitudes at six loops!





This work is a small step in the broader program of exploring the hidden permutation symmetry of ABJM squared amplitudes and possible relations with other quantities such as correlators and Wilson loops, as well as underlying positive geometries such as the ABJM amplituhedron~\cite{He:2021llb,Huang:toa,Lukowski:2021fkf,Oren-Perlstein:2025ljb,He:2022cup,He:2023rou,Lukowski:2023nnf,Ferro:2025pij}. For four points, it would be interesting to reorganize the entire logarithm of the amplitude in terms of bipartite negative geometries and fix the canonical forms for such geometries with different ``topologies". An important physical application of such integrands is that by integrating them we can extract the important quantity known as the cusp anomalous dimensions ~\cite{He:2023exb,Henn:2023pkc,Li:2024lbw,Lagares:2024epo}, and how to do this at $L=6$ remains an important open problem. It would also be highly desirable to understand how vanishing cuts in loop sub-topologies connect to the conjecture~\cite{Li:2024lbw} that loop topologies do not contribute to $\Gamma_{\text{cusp}}$. Along the same line, it would also be very interesting to study periods of these weight-$3$ $f$-graphs and their relations with certain ``integrated correlators" similar to the SYM case~\cite{Wen:2022oky, Brown:2023zbr, Zhang:2024ypu, He:2025vqt, He:2025lzd}, and integrations for higher-point amplitudes in ABJM theory. 

Another intriguing direction is to study the squared amplituhedron and correlahedron in $D=3$ via dimension reduction of their four-dimensional counterparts~\cite{Eden:2017fow,Dian:2021idl,He:2024xed,He:2025rza}. While it works nicely for the amplituhedra, directly reducing the correlahedron in $D=4$ by flipping all signs except for $\Delta^2>0$ does not seem to work: the one-loop object does not vanish, and it produces a two-loop correlator in \eqref{eq:cor42} with additional kissing-triangles as the ``chamber form". Despite these discouraging results, it produces correct lightlike limits, {\it i.e.} vanishes at four-point one loop as well as five-point tree-level and also gives correct squared amplitude at four-point two-loop and six-point tree-level.  Moreover, the geometry after taking the lightlike limit can be directly defined: the four-point squared ABJM amplituhedron in momentum twistor space can be defined as
\begin{equation}\label{eq:D32}
   \begin{aligned}
       &\langle1234\rangle<0, \frac{\langle A_j B_j12\rangle}{\langle A_j B_j 41\rangle}<0,\frac{\langle A_j B_j23\rangle}{\langle A_j B_j 41\rangle}<0,\frac{\langle A_j B_j34\rangle}{\langle A_j B_j 41\rangle}<0,\frac{\langle A_iB_iA_jB_j\rangle}{\langle A_i B_i41\rangle\langle A_j B_j41\rangle}<0\\
       &\text{with} A_i\cdot\Omega\cdot B_i=0,\quad i,j=1,2,\ldots,L.
   \end{aligned}
\end{equation}
and summing over all $2^L$ sign choices reproduces exactly the combination appearing in~\eqref{eq:sqdef} at all loop orders. We leave deeper exploration for future work.

For $n\geq 8$ cases, it is already interesting to study tree-level amplitudes from the $f$-graphs, where \eqref{eq:sqdet} provides a powerful tool. Regarding this, we point out that the determinant formula is general and it would be interesting to apply it to other theories, such as squaring super-amplitudes in SYM and supergravity in four dimensions~\cite{Chicherin:2025keq}.
Moreover, while our computation in this paper is formulated in spinor-helicity variables, it would also be very interesting to extend this determinant formula to momentum twistor space~\cite{ArkaniHamed:2009vw, Arkani-Hamed:2012zlh,Elvang:2014fja, He:2023rou}, where (super)momentum conservation is manifest. Such an extension could be crucial for a better understanding of underlying positive geometries, including the ABJM amplituhedra in both spaces, the square of such objects and analogous ``correlahedron" as their off-shell generalizations.

Last but not least, it would be extremely interesting if our study could shed more lights into the origin of this hidden permutation symmetry, and the longstanding problem of amplitude/Wilson loop/correlator triality in ABJM theory.

\begin{CJK*}{UTF8}{}
\CJKfamily{gbsn}
\acknowledgments
It is our pleasure to thank Chia-Kai Kuo, Xichen Li, Canxin Shi and Yichao Tang for helpful discussions and/or collaborations on related topics. SH has been supported by the National Natural
Science Foundation of China under Grant No. 12225510,
12447101, 12247103, and by the New Cornerstone Science
Foundation. YZ was also supported in part by the Deutsche Forschungsgemeinschaft (DFG, German Research Foundation) Projektnummer 508889767/FOR5582 “Modern Foundations of Scattering
Amplitudes”.
\end{CJK*}

\appendix
\section{The definition of three-loop planar DCI integrands}\label{app:definition}
Here we present the definition of DCI integrals at $L=3$ extracted from the planar $f$-graph in ABJM theory. All the $x_{13}^{2},x_{24}^2$ are omitted and those valency-$4$ vertices $i$ has a numerator $\epsilon_i$. The entire integrand is also included in the ancillary file.
\begin{align*}
&\includegraphics[align=c,scale=0.2]{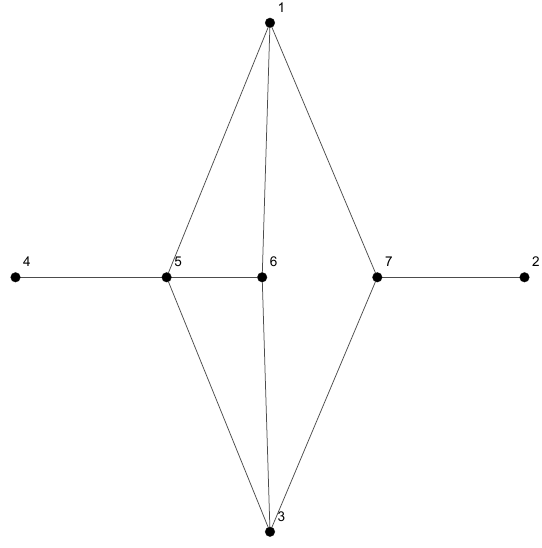}=\frac{\epsilon_5}{x_{1,5}^2 x_{1,6}^2 x_{1,7}^2 x_{2,7}^2 x_{3,5}^2 x_{3,6}^2 x_{3,7}^2 x_{4,5}^2 x_{5,6}^2},\includegraphics[align=c,scale=0.2]{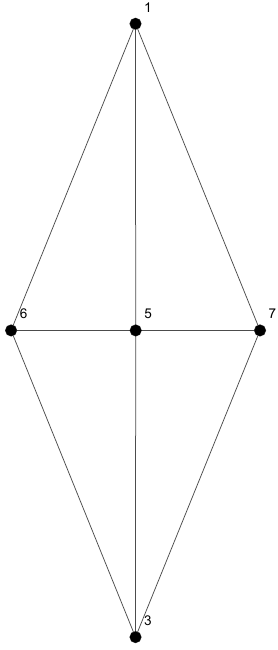}=\frac{\epsilon_5}{x_{1,5}^2 x_{1,6}^2 x_{1,7}^2 x_{3,5}^2 x_{3,6}^2 x_{3,7}^2 x_{5,6}^2 x_{5,7}^2}\\
&\includegraphics[align=c,scale=0.2]{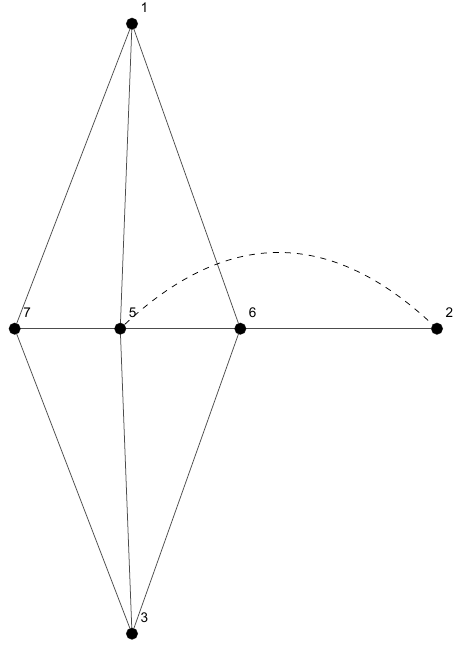}=\frac{\epsilon_6x_{2,5}^2}{x_{1,5}^2 x_{1,6}^2 x_{1,7}^2 x_{2,6}^2 x_{3,5}^2 x_{3,6}^2 x_{3,7}^2 x_{5,6}^2 x_{5,7}^2},\includegraphics[align=c,scale=0.2]{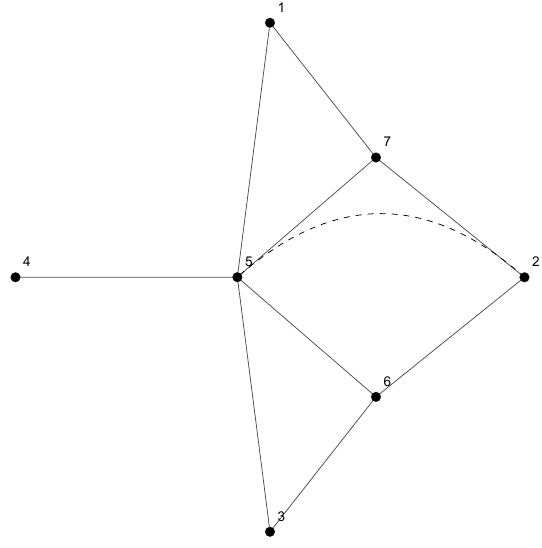}=\frac{\epsilon_5x_{2,5}^2}{x_{1,5}^2 x_{1,7}^2 x_{2,6}^2 x_{2,7}^2 x_{3,5}^2 x_{3,6}^2 x_{4,5}^2 x_{5,6}^2 x_{5,7}^2}\\
&\includegraphics[align=c,scale=0.2]{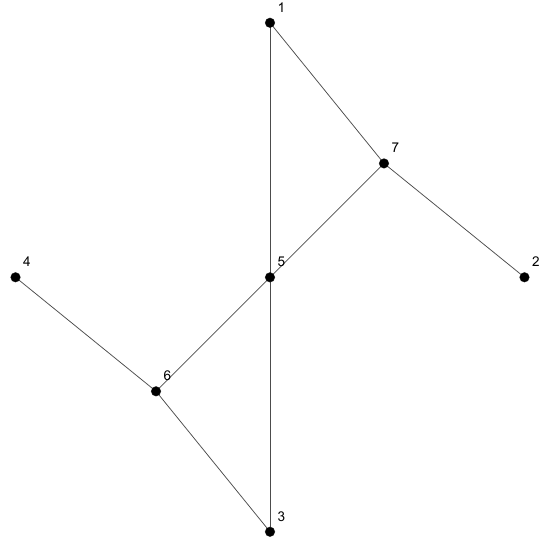}=\frac{\epsilon_5}{x_{1,5}^2 x_{1,7}^2 x_{2,7}^2 x_{3,5}^2 x_{3,6}^2 x_{4,6}^2 x_{5,6}^2 x_{5,7}^2},\includegraphics[align=c,scale=0.2]{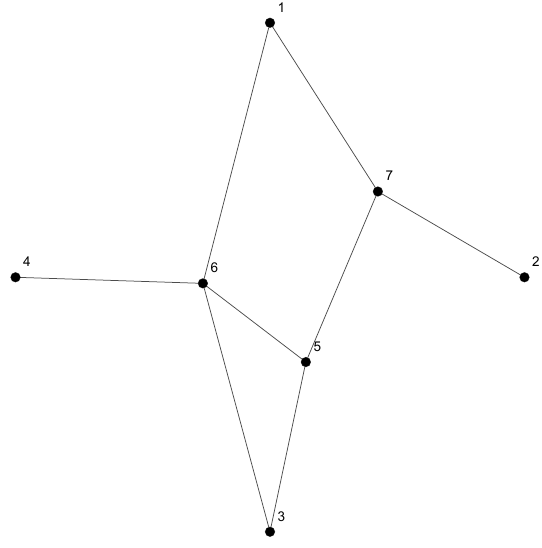}=\frac{\epsilon_6}{x_{1,6}^2 x_{1,7}^2 x_{2,7}^2 x_{3,5}^2 x_{3,6}^2 x_{4,6}^2 x_{5,6}^2 x_{5,7}^2}\\
&\includegraphics[align=c,scale=0.2]{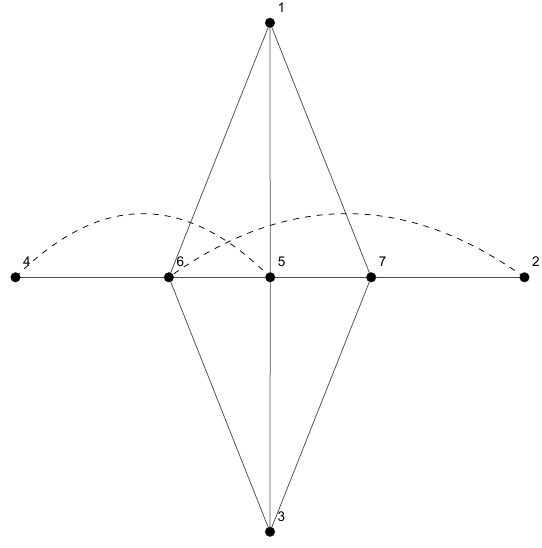}=\frac{\epsilon_7x_{2,6}^2 x_{4,5}^2}{x_{1,5}^2 x_{1,6}^2 x_{1,7}^2 x_{2,7}^2 x_{3,5}^2 x_{3,6}^2 x_{3,7}^2 x_{4,6}^2 x_{5,6}^2 x_{5,7}^2},\includegraphics[align=c,scale=0.2]{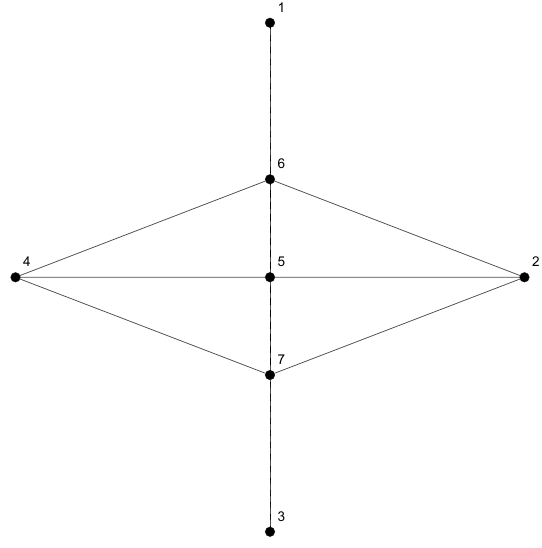}=\frac{\epsilon_5x_{1,7}^2 x_{3,6}^2}{x_{1,6}^2 x_{2,5}^2 x_{2,6}^2 x_{2,7}^2 x_{3,7}^2 x_{4,5}^2 x_{4,6}^2 x_{4,7}^2 x_{5,6}^2 x_{5,7}^2}\\
&\includegraphics[align=c,scale=0.2]{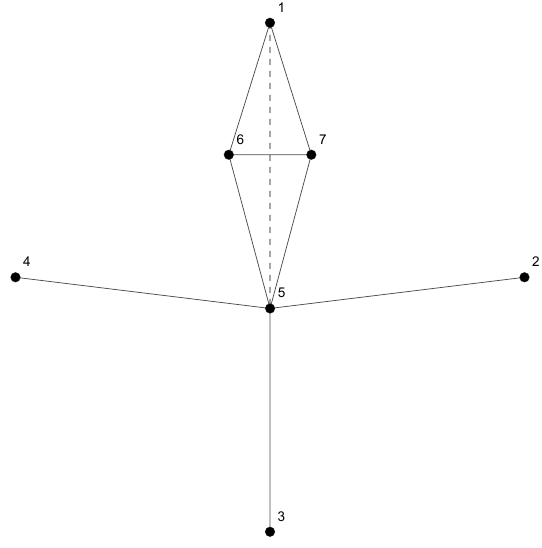}=\frac{\epsilon_5x_{1,5}^2}{x_{1,6}^2 x_{1,7}^2 x_{2,5}^2 x_{3,5}^2 x_{4,5}^2 x_{5,6}^2 x_{5,7}^2 x_{6,7}^2},\includegraphics[align=c,scale=0.2]{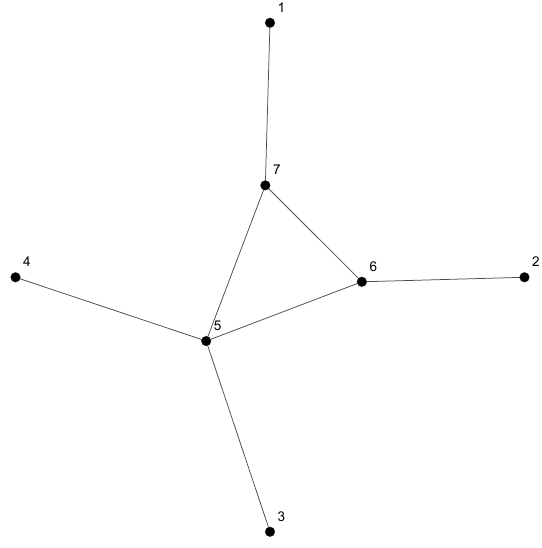}=\frac{\epsilon_5}{x_{1,7}^2 x_{2,6}^2 x_{3,5}^2 x_{4,5}^2 x_{5,6}^2 x_{5,7}^2 x_{6,7}^2}\\
&\includegraphics[align=c,scale=0.2]{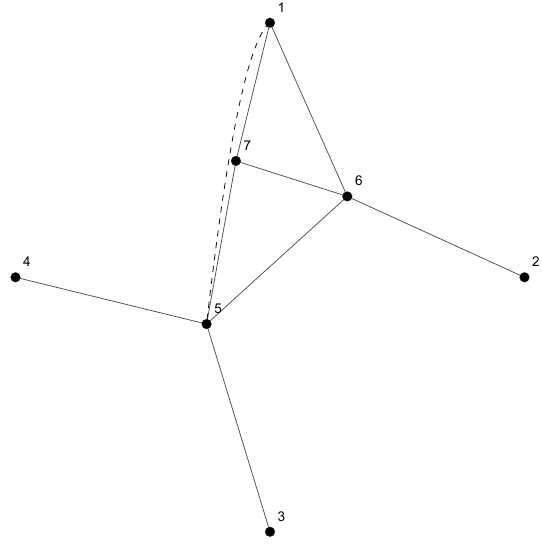}=\frac{\epsilon_6x_{1,5}^2}{x_{1,6}^2 x_{1,7}^2 x_{2,6}^2 x_{3,5}^2 x_{4,5}^2 x_{5,6}^2 x_{5,7}^2 x_{6,7}^2},\includegraphics[align=c,scale=0.2]{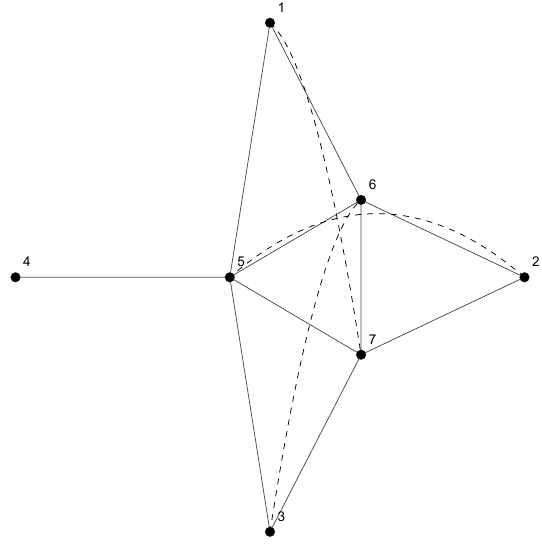}=\frac{\epsilon_5x_{1,7}^2 x_{2,5}^2 x_{3,6}^2}{x_{1,5}^2 x_{1,6}^2 x_{2,6}^2 x_{2,7}^2 x_{3,5}^2 x_{3,7}^2 x_{4,5}^2 x_{5,6}^2 x_{5,7}^2 x_{6,7}^2}\\
&\includegraphics[align=c,scale=0.2]{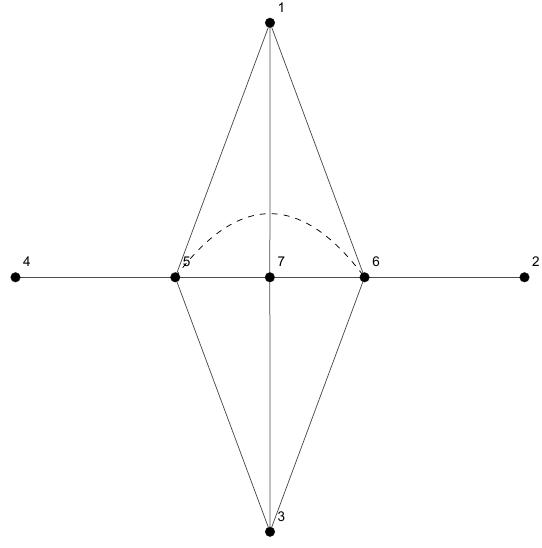}=\frac{\epsilon_7x_{5,6}^2}{x_{1,5}^2 x_{1,6}^2 x_{1,7}^2 x_{2,6}^2 x_{3,5}^2 x_{3,6}^2 x_{3,7}^2 x_{4,5}^2 x_{5,7}^2 x_{6,7}^2}
\end{align*}

\section{Proof of Theorem \eqref{eq:vanishLS}}\label{app:proofvanish}

Let
$
V=\left.\hat{C}\right|_{\text{sol}}$ and $
W=\left.\hat{C}^{\prime}\right|_{\text{sol}^{\prime}},
$
with branch signs $\sigma(V),\sigma(W)\in\{\pm1\}$.
The row vectors of $V$ and $W$ are denoted by $v_i$ and $w_j$ $(i,j=1,\ldots,k)$ . We define a $k\times k$ matrix $P$ by
\begin{equation}
P := V\,\eta\, W^{T}, \qquad
P_{ij} = (-1)^j\, v_i w_j^{T}.
\end{equation}

We shall prove that, for $V\neq W$,
\begin{equation}
    \sigma(V)\sigma(W)=(-1)^{k+1}\Rightarrow \det P=0.
\end{equation}

\begin{proof}
We begin by permuting the columns of $V$ and $W$ by an invertible permutation matrix $P_\pi$ so that
\begin{equation}
\eta' := P_\pi^T \eta P_\pi = \mathrm{diag}(\underbrace{1,\ldots,1}_{k},\underbrace{-1,\ldots,-1}_{k}).
\end{equation}
Define $V' := V P_\pi$ and $W' := W P_\pi$.
Since $P_\pi$ is invertible, $V'$ and $W'$ are also elements of the orthogonal Grassmannian $OG(k,2k)$ with respect to the metric $\eta'$, and their branch are unchanged.

Using the $\mathrm{GL}(k)$ gauge-fixing, we may further bring $V'$ and $W'$ to the standard forms
\begin{equation}
V' = [\,I_k\ \ C\,], \qquad
W' = [\,I_k\ \ D\,],
\end{equation}
where $C,D\in O(k)$ are orthogonal matrices.
In this parametrization, the branch signs are given by
\begin{equation}
\sigma(V) = \det C, \qquad
\sigma(W) = \det D .
\end{equation}

The matrix $P$ is similar to $P^\prime$
\begin{equation}
P\sim P^\prime = V^\prime\eta^\prime W^{\prime T}
   = (I_k - C D^T)
\end{equation}
Taking determinants yields
\begin{equation}
\det P=\det P^\prime
= \det(I_k - C D^T)
\equiv 
\det(I_k - Q),
\end{equation}
where we have defined
$
Q := C D^T \in O(k).
$
Therefore, $P$ is full rank if and only if
$
\det(I_k - Q)\neq 0
$. Since $Q$ is an orthogonal matrix,
\begin{equation}
\det(I_k - Q)
= \det(Q)\,\det(Q^T - I_k)
= (-1)^k \det(Q)\,\det(I_k - Q).
\end{equation}
It follows that
\begin{equation}
\bigl(1 - (-1)^k \det Q\bigr)\,\det(I_k - Q) = 0.
\end{equation}
Hence if
\begin{equation}
\det Q = (-1)^{k+1}\neq(-1)^k,
\end{equation}
$\det(I_k - Q)$ must vanish.

Finally, using $\det Q = \det C\,\det D = \sigma(V)\sigma(W)$, we conclude that
\[
\sigma(V)\sigma(W) = (-1)^{k+1}\Rightarrow\det P=0.
\]
Equivalently, when $k$ is even, products between different branches will vanish, while at odd $k$, products between the same branches vanish.

\end{proof}

\bibliographystyle{JHEP}
\bibliography{amplitudes_refs.bib}

\end{document}